# Inelastic neutron scattering and lattice dynamical calculations in $BaFe_2As_2$

R. Mittal[1], Y. Su[1], S. Rols[2], T. Chatterji[3], S. L. Chaplot[4], H. Schober[2], M. Rotter[5], D. Johrendt[5] and Th. Brueckel[1,6]

[1]*Juelich Centre for Neutron Science, IFF, Forschungszentrum Juelich, Outstation at FRM II, Lichtenbergstr. 1, D-85747 Garching, Germany*
[2]*Institut Laue-Langevin, BP 156, 38042 Grenoble Cedex 9, France,*
[3]*Juelich Centre for Neutron Science, Forschungszentrum Juelich, Outstation at Institut Laue-Langevin, BP 156, 38042 Grenoble Cedex 9, France*
[4]*Solid State Physics Division, Bhabha Atomic Research Centre, Trombay, Mumbai 400 085, India*
[5]*Department Chemie und Biochemie, Ludwig-Maximilians-Universitaet Muenchen, Butenandtstrasse 5-13 (Haus D), D-81377 Muenchen, Germany*
[6]*Institut fuer Festkoerperforschung, Forschungszentrum Juelich, D-52425 Juelich, Germany*

We report here first extensive measurements of the temperature dependence of phonon density of states of $BaFe_2As_2$, the parent compound of the newly discovered FeAs-based superconductors, using inelastic neutron scattering. The experiments were carried out on the thermal time-of-flight neutron spectrometer IN4 at the ILL on a polycrystalline sample. There is no appreciable change in the spectra between T = 10 K and 200 K, although the sample undergoes a magnetic as well as a tetragonal-to-orthorhombic structural phase transition at 140 K. This indicates a rather harmonic phonon system. Shell model lattice dynamical calculations based on interatomic potentials are carried out to characterize the phonon data. The calculations predict a shift of the Ba-phonons to higher energies at 4 GPa. The average energy of the phonons of the Ba-sublattice is also predicted to increase on partial substitution of Ba by K to $Ba_{0.6}K_{0.4}$. The calculations show good agreement with the experimental phonon spectra, and also with the specific heat data from the literature.

The discovery of high transition temperature ($T_c$) superconductivity in fluorine-doped RFeAsO (R = rare earth) and K-doped $BaFe_2As_2$ has attracted immense attention [1-21] in the scientific community. It is important to note that these compounds have high superconducting transition temperatures without requiring the presence of copper oxide layers. The application of hydrostatic pressure in F-doped LaFeAsO results [4] in an increase in $T_c$ from 25 K to 43 K. Similarly the substitution of rare earth atoms equally results [5] in an increase of Tc to 43 K at ambient pressure in $Sm_{1-x}F_xFeAsO$. A $T_c$ of 38 K has been reported [7] in bilayer compound $Ba_{0.60}K_{0.40}Fe_2As_2$. Pressure-induced superconductivity has been reported [12,13] in $MFe_2As_2$ (M=Ba, Ca and Sr) These measurements indicate that a $T_c$ of 12 K, 27 K and 29 K can be achieved at high pressures in the parent compounds $CaFe_2As_2$, $SrFe_2As_2$ and $BaFe_2As_2$, respectively. Anomalies in the specific heat, electrical resistance and magnetic susceptibility indicate [7] structural and magnetic phase transition in $BaFe_2As_2$ at 140 K. X-ray diffraction measurements indicate that an orthorhombic (space group *Fmmm*) to tetragonal (*I4/mmm*) structural phase transition takes place at 140 K [7]. The magnetic ordering in this compound has been confirmed by neutron diffraction on both powder [19] and single-crystal samples [18]. Theoretical electronic structure calculations [8,9,20] have been reported for LaFeAsO and $BaFe_2As_2$. These calculations show that in LaFeAsO electronic bands around the Fermi level are formed mainly by Fe-As layers, whereas the bands of La-O layers are far from the Fermi level. Therefore superconductivity in these compounds is believed to be mainly due to the structural and electronic states of the Fe-As layers. Structural, electronic and magnetic properties of $BaFe_2As_2$ and LaFeAsO are remarkably similar, which makes $BaFe_2As_2$ a parent compound for oxygen-free superconductors.

Concerning the dynamics Raman [10] and infrared data [21] have been reported in LaFeAsO, while Raman modes have been experimentally measured [14] in $SrFe_2As_2$. Ab-initio calculation [15] of the phonon spectra have been reported in LaFeAsO. The zone-centre phonon modes have been calculated [14] for $SrFe_2As_2$ using shell model lattice dynamics. To our knowledge, the measurements of the phonon density of states are restricted to superconducting $LaO_{0.87}F_{0.13}FeAs$ [16] below 20 meV. A strong electron-phonon coupling of in plane Fe breathing modes has been suggested for the $LaO_{1-x}F_xFeAs$ superconductor [17]. Meanwhile, a magnetic instability, due to either magnetic ordering of a localized iron magnetic moment or spin-density waves in a more itinerant picture has been experimentally realized in the parent compounds [2]. This raises possibility that superconductivity in iron pnictides might also be mediated by spin fluctuations, in a case similar to cuprates. The mechanism of superconductivity, and in particular the role of lattice dynamics in superconducting pair formation in these newly discovered materials that superconduct with relatively high $T_c$, although not as high as that found in cuprates, is still to be settled. Meanwhile it is necessary to study phonon dynamics carefully in these materials. This has motivated us to carry out measurements of the temperature dependence of the phonon density of states in $BaFe_2As_2$. The phonons spectra as well as some thermodynamic properties have equally been calculated.

The polycrystalline sample of $BaFe_2As_2$ was prepared and characterized as reported elsewhere [7]. The inelastic neutron scattering experiments were performed on a polycrystalline sample of $BaFe_2As_2$ using the thermal time-of-flight neutron spectrometer IN4 at the Institut Laue Langevin (ILL), France. The incident neutron wavelength of 1.18 Å was chosen which allowed the spectra to be obtained in the neutron energy loss mode (Stoke side) up to 45 meV. The detector bank covered scattering angles up to 120°. About 1.6 g of polycrystalline $BaFe_2As_2$ sample was placed inside a thin aluminum foil. The temperature dependent data were taken at 10 K, 100 K and 200K. The data were processed in order to obtain the dynamical structure factor $S(Q, E)$ and the neutron weighted "generalized" phonon density of states $g^{(n)}(E)$. The so called "incoherent approximation" is used to derive $g^{(n)}(E)$ which implies averaging the spectra over the whole scattering angle [22]. In the energy loss mode, the neutron-weighted phonon density of states are obtained [23] from the measured scattering function $S(Q,E)$ via

$$g^{(n)}(E) = A < \frac{e^{2W_k(Q)}}{Q^2} \frac{E}{n(E,T)+1} S(Q,E) > \qquad (1)$$

$$g^n(E) = B \sum_k \{\frac{4\pi b_k^2}{m_k}\} g_k(E) \qquad (2)$$

where $n(E,T) = [\exp(E/k_BT)-1]^{-1}$, $A$ and $B$ are normalization constants and $b_k$, $M_k$, and $g_k(E)$ are, respectively, the neutron scattering length, mass, and partial density of states of the $p$-th atom in the unit cell. The quantity within < ---- > represents the proper average over all $Q$ values at a given energy. $2W(Q)$ is the Debye-Waller factor.

The present lattice dynamical calculations are carried out using a shell model. Each ion consists of a core and an adiabatic, i.e. mass less shell of charge, which are connected by a harmonic spring constant. The shell can displace itself from the core causing a dipole, leading to a proper description of the crystal dielectric behavior. The form of the potentials used [24,25] for the calculations is:

$$V(r) = \{\frac{e^2}{4\pi\varepsilon_o}\}\{\frac{Z(k)Z(k')}{r}\} + a\exp\{\frac{-br}{R(k)+R(k')}\} ,  \qquad (3)$$

where r is the separation between the atoms of type k and k', Z(k) and R(k) are empirical charge and radius parameters of the atom of type k, $1/(4\pi\varepsilon_o) = 9 \times 10^9$ Nm$^2$/Coul$^2$, a=1822 eV, and b=12.364.

The covalent nature of Fe-As bond is described by the following potential

$$V(r) = -D\exp[-n(r-r_o)^2/(2r)] . \qquad (4)$$

The radii parameters used in our calculations are R(Ba) = 2.20 Å, R(Fe) = 0.30 Å and R(As) =2.48 Å. Partial charges of Z(Ba) = 1.50, Z(Fe) = 0.75 and Z(As) = -1.50 are used in the calculations. The parameters of the stretching potential are D=2.4 eV, n=7 Å$^{-1}$, r$_o$ =2.392 Å. The shell charge Y(As) and shell-core force constant K(As) for As atoms are -1.0 and 140 eV/Å$^2$. The parameters of the potential are determined from the static and dynamic equilibrium conditions of the crystal lattice. The lattice dynamical calculations are carried out using DISPR [26] software package developed at Trombay.

The phonon spectra collected for BaFe$_2$As$_2$ at 10 K, 100 K and 200 K are shown in Figs. 1. BaFe$_2$As$_2$ has an orthorhombic-to-tetragonal structural and magnetic phase transition at 140 K when increasing the temperature. The orthorhombic distortion is rather small. The experimental phonon spectrum at 200 K in the tetragonal phase is quite similar to those in the orthorhombic phase at 10 K and 100 K. The experimental S(Q,E) plots (Fig. 2) at 10 K, 100 K and 200 K show no immediately recognizable signs of magnetic excitations in BaFe$_2$As$_2$ in our measurements between 5 and 45 meV. Our present interatomic potential produces a stable tetragonal structure. The calculations are compared (Fig. 1) with the experimental phonon data at 10 K. The calculations compare very well with the experimental data except for a peak at about 21.5 meV, which is not reproduced by the calculations. The intensity of this peak is observed to scale with the Bose factor at all the temperatures. The shape of this peak shows an identifiable temperature variation between 10 K and 200 K. The experimental S(Q,E) (Fig. 2) plots show that this peak does not arise from intensity at low Q. This suggests that the origin of this peak is not magnetic.

In order to understand the contribution of various atomic motions to the phonon spectra we have calculated the partial densities of states (Fig. 3). The Ba atoms mainly contribute in the 0–20 meV range, while the Fe and As atoms contribute in the entire 0–40 meV ranges, respectively. Above 30 meV the contributions are mainly due to Fe-As stretching modes.

Application of hydrostatic pressure up to 4 GPa in F-doped LaFeAsO results [4] in an increase of Tc from 25 K to 43 K. Similarly, the parent non-superconducting $BaFe_2As_2$ compound becomes superconducting [13] with a $T_c$ of 27 K at 3.8 GPa. In order to understand the role of phonon in increasing $T_c$ we have calculated phonon spectra at 4 GPa. We find that the maximum shift with increasing pressure is encountered for the Ba phonons (Fig. 3). At ambient pressure $BaFe_2As_2$ shows superconductivity if we partially substitute Ba with K. Pressure generates nearly the same effect as the partial substitution of Ba by K atoms. In order to understand this we have carried out super cell calculations (3 × 3 × 3) for $Ba_{2/3}K_{1/3}Fe_2As_2$. The calculated phonon spectra for $Ba_{2/3}K_{1/3}Fe_2As_2$ are also shown in Fig. 3. The atomic mass of K (39.10 amu) is smaller in comparison to Ba (137.34 amu). This results in an increase of the average phonon energy of $Ba_{0.6}K_{0.4}$ site atoms only. Experimental measurements of Raman modes in $SrFe_2As_2$ also show [14] that substitution of K for Sr has little effect on the frequencies of modes involving As and Fe atoms. Modes involving Sr or $Sr_{0.6}K_{0.4}$ could not be measured by Raman spectroscopy since these $A_{2u}$ or $E_u$ modes are only sensitive to the infra red technique.

The calculated one-phonon density of states $g(E)$ is shown in Fig. 3, and is used to compute the specific heat $C_V(T)$ as a function of temperature. The difference $C_P$-$C_V$= $\alpha_V^2$ $BVT$, where $\alpha_V$ is the volume thermal expansion and $B$ is the bulk modulus. For comparison with the experimental specific heat of $BaFe_2As_2$ the experimental electronic specific heat coefficient $\gamma$(=23 mJoule/moleK$^4$) [11] is added to the calculated phonon contribution. The calculated $C_P(T)$ is compared with the experimental data [11] in Fig. 4(a). For BaFe$_2$As$_2$ the calculated value of bulk modulus is 69 GPa. Experimental data is not available for comparison.

In conclusion, we have measured the temperature dependence of the phonon density of states in the parent compound BaFe$_2$As$_2$. We find no indications for strong anharmonic effects. Lattice dynamical calculations are carried out for microscopic understanding of the phonon spectra. The comparison between our experimental phonon data and shell model is very good apart from one peak in the experimental spectra that could not be reproduced by the calculations. We have shown that the effect of pressure on the phonon vibrations is the strongest for the barium modes and so also is the effect of a partial substitution of Ba by K atoms on the modes of the Ba-K sublattice.

FIG. 1. (Color online) Comparison between the calculated and experimental phonon spectra of BaFe$_2$As$_2$. For better visibility the experimental phonon spectra at 100 K and 200 K are shifted along the y-axis by 0.04 meV$^{-1}$ and 0.08 meV$^{-1}$, respectively. The calculated spectra have been convoluted with a Gaussian of FWHM of 3 meV in order to describe the effect of energy resolution in the experiment.

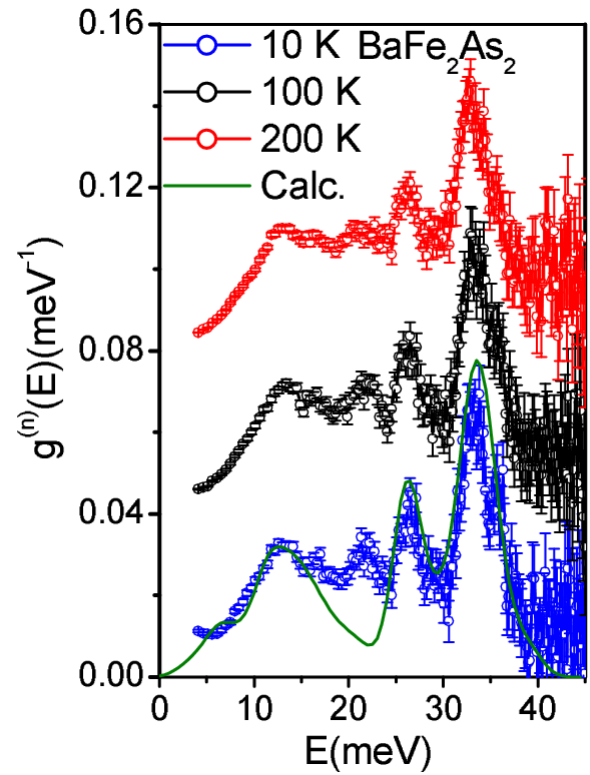

FIG. 2. (Color online) The experimental S(Q,E) plots for $BaFe_2As_2$ at 10 K, 100 K and 200 K.

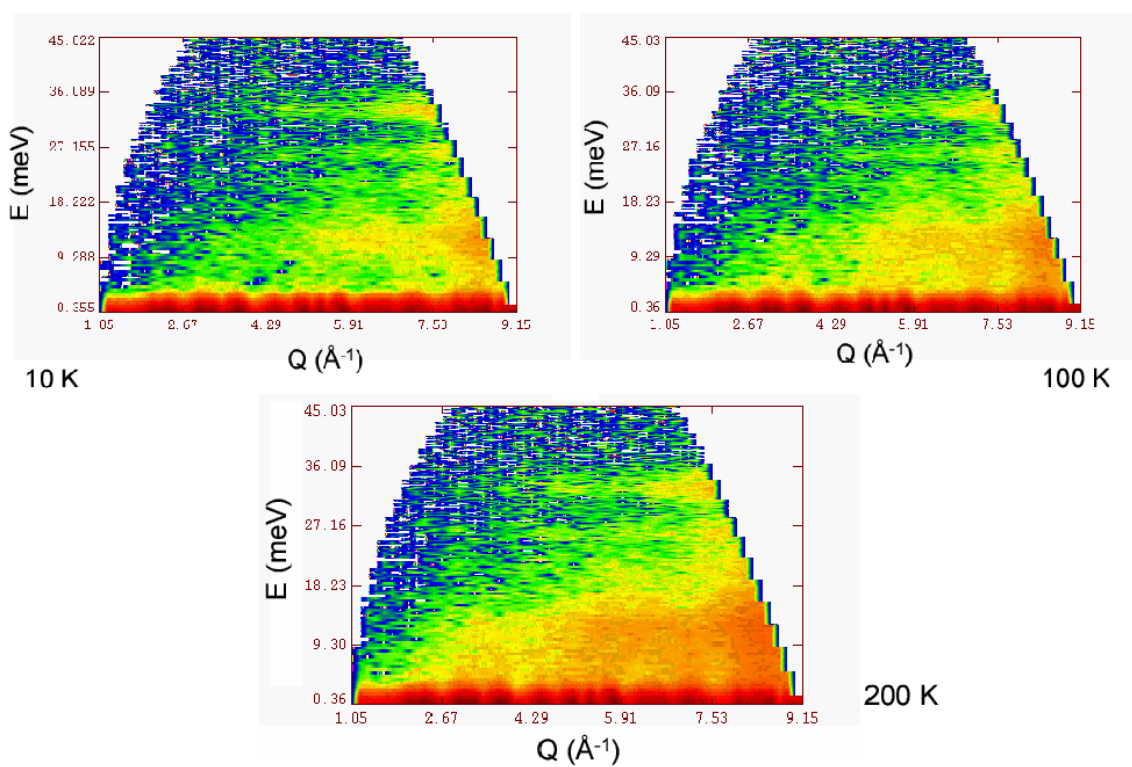

FIG. 3. (Color online) The calculated partial density of states of the various atoms in $BaFe_2As_2$ and $Ba_{2/3}K_{1/3}Fe_2As_2$.

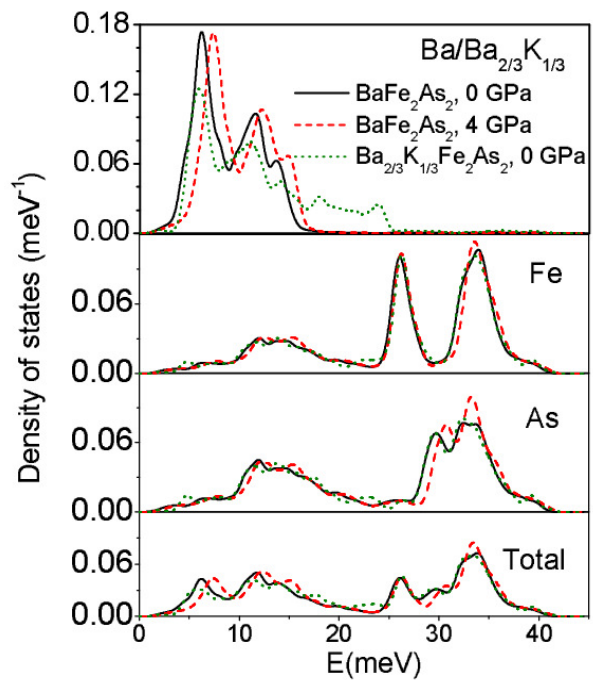

FIG. 4. Comparison between the calculated and experimental specific heat [11].

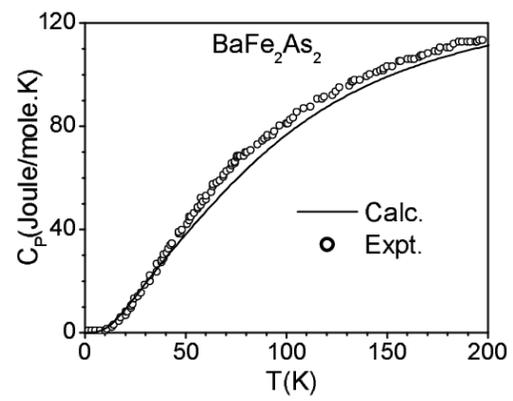